\documentclass[a4paper,11pt]{article}
\pdfoutput=1 

\usepackage{jinstpub} 

\title{\boldmath Characterization of the demonstrator of the fast silicon monolithic ASIC for the TT-PET project}

\author[a,1]{L. Paolozzi,\note{Corresponding author.}}
\author[b]{Y. Bandi,}
\author[c]{R. Cardarelli,}
\author[a]{S. D\'ebieux,}
\author[a]{Y. Favre,}
\author[a]{D. Ferr\`ere,}
\author[b]{D. Forshaw,}
\author[a,d]{D. Hayakawa,}
\author[a]{G. Iacobucci,}
\author[e]{M. Kaynak,}
\author[b]{A. Miucci,}
\author[a,f]{M. Nessi,}
\author[a,d]{E. Ripiccini,}
\author[e]{H. R\"ucker,}
\author[a]{P. Valerio,}
\author[b]{and M. Weber}

\affiliation[a]{University of Geneva,\\ Rue du G\'en\'eral-Dufour 24, Geneva, Switzerland}
\affiliation[b]{University of Bern,\\ Sidlerstrasse 5, Bern, Switzerland}
\affiliation[c]{INFN Section of Roma Tor Vergata,\\ Via della ricerca scientifica 1, Roma, Italy}
\affiliation[d]{Institute of Translational Molucular Imaging (ITMI), University of Geneva\\ Geneva, Switzerland}
\affiliation[e]{IHP - Leibniz-Institut f\"ur innovative Mikroelektronik \\ Im Technologiepark 25, Frankfurt (Oder), Germany}
\affiliation[f]{CERN,\\ Geneva, Switzerland}

\emailAdd{lorenzo.paolozzi@unige.ch}

\abstract{The TT-PET collaboration is developing a small animal TOF-PET scanner based on monolithic silicon pixel sensors in SiGe BiCMOS technology. The demonstrator chip, a small-scale version of the final detector ASIC, consists of a $ 3 \times 10 $ pixel matrix integrated with the front-end, a $ 50 ~\mathrm{ps} $ binning TDC and read out logic. The chip, thinned down to $ 100 ~\mathrm{\mu m} $ and backside metallized, was operated at a voltage of $ 180 ~\mathrm{V} $. The tests on a beam line of minimum ionizing particles show a detection efficiency greater than $ 99.9 ~\mathrm{ \% } $ and a time resolution down to $ 110 ~\mathrm{ps} $.}


\begin{document}
\maketitle
\flushbottom

\section{The TT-PET ASIC}
\label{sec:intro}
\subsection{Challenges and previous results}

The development of a monolithic silicon pixel detector with $ 30 ~\mathrm{ps ~RMS} $ ($ 70 ~\mathrm{ps~FWHM} $) time resolution for $ 511 ~\mathrm{keV} $ photons is the main challenge of the Thin-TOF PET (TT-PET) \cite{ttpet} scanner, a small-animal PET system that makes use of a stack of silicon sensors to detect electrons from converted photons. The strategy adopted for this chip, already described in \cite{testbeam_2017}, is to integrate the sensor and the front-end electronics in a SiGe BiCMOS process, to produce an ultra-fast, low-power silicon pixel detector. The ASIC will have a thickness of $ 100 ~\mathrm{\mu m} $, comprising the BiCMOS processing and a depletion depth of $ 80 ~\mathrm{\mu m} $ on a high resistivity substrate. In order to saturate the drift velocity of the charge carriers and provide a very uniform weighting field, the sensor will be backside metallized and will operate with a high voltage of approximately $ 200 ~\mathrm{V} $. The large number of detection elements in a small animal scanner (1920 monolithic chips) sets a constraint on the power consumption to $ 200 ~\mathrm{\mu W/channel} $ for $ 500 \times 500 ~\mathrm{~\mu m ^2} $ pixels.

The target of $ 30 ~\mathrm{ps ~RMS} $ time resolution for electrons from the conversion of $ 511 ~\mathrm{keV} $ photons corresponds to approximately $ 100 ~\mathrm{ps ~RMS} $ for perpendicularly incident minimum ionising particles (MIPs), due to the smaller charge that MIPs generate into the sensor \cite{testbeam_2017}. The first TT-PET ASIC prototype achieved a record time resolution for a monolithic pixel sensor of $ 220 ~\mathrm{ps ~RMS} $ with MIPs, with a power consumption of $ 350 ~\mathrm{\mu W} $ for $ 900 \times 450 ~\mathrm{\mu m ^2} $ pixels \cite{testbeam_2017}. That chip was operated on a $ 700 ~\mathrm{\mu m} $ thick, high resistivity substrate, with a depletion depth of $ 150 ~\mathrm{\mu m} $. An improvement of a further factor two from this result is possible by reducing the sensor thickness at fixed detector capacitance and metallizing the backside of the chip, which will increase the uniformity of the weighting field for the induced current and saturate the drift velocity of the charge carriers \cite{silicon_tres, 100ps}.

\subsection{The demonstrator}

\begin{figure}[htpb]
\centering
\includegraphics[width=.95\textwidth,trim=0 0 0 0]{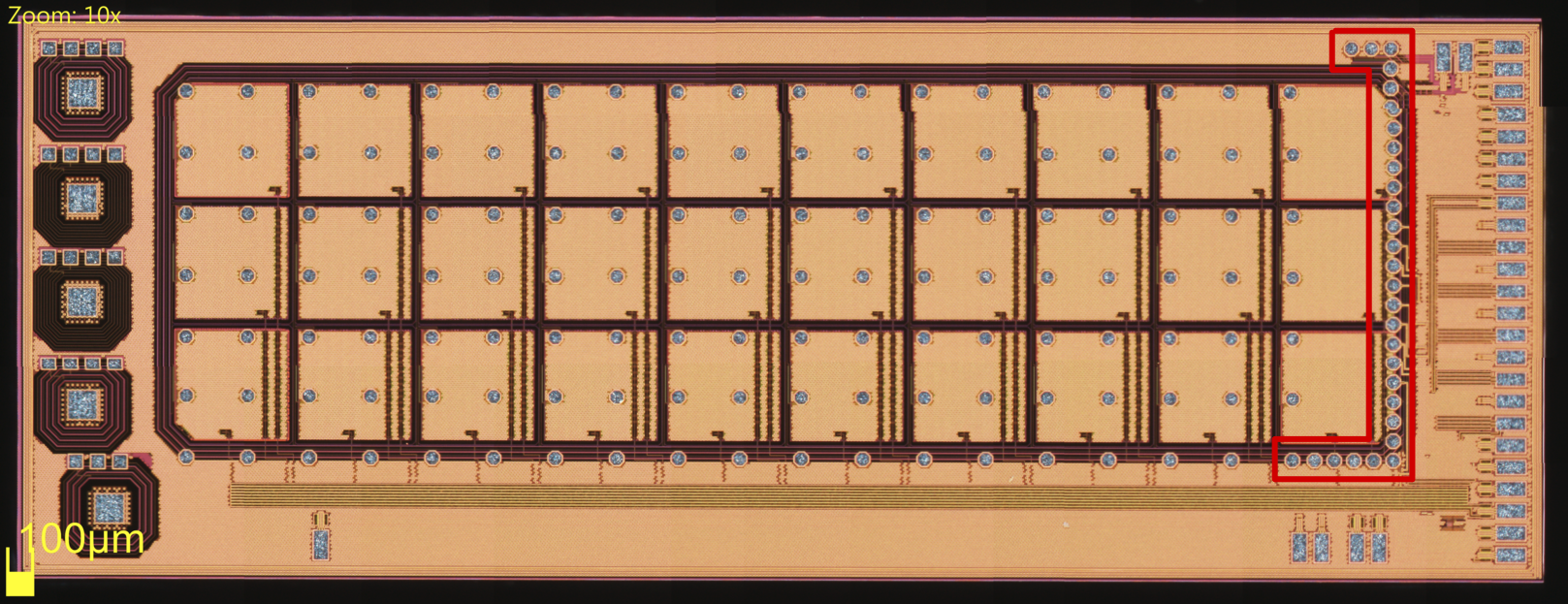}
\caption{\label{fig:p0_layout} Microscope picture of the demonstrator of the fast monolithic silicon pixel detector of the TT-PET project. The pixel matrix consists of three rows of 10 pixels each. The guard ring is partially visible from the opening of the metallization around the pixel matrix. The signals from the pixels are routed to the front-end, distributed over the long side of the chip, outside the guard ring. The TDC, logic and I/O is in the right periphery of the chip. The rectangular wire-bonding pads are connected to smaller octagonal bump-bonding pads, sitting close to the edge of the rightmost pixel column (inside the area identified by the red lines). The five structures on the left are independent guard-ring test structures.}
\end{figure}

The ASIC prototype described in this work, and hereby called demonstrator \cite{pierpaolo_chip}, is a monolithic chip developed for the TT-PET project in the SiGe BiCMOS process SG13S from IHP microelectronics \cite{sg13s}. The demonstrator, shown in Figure~\ref{fig:p0_layout}, was designed to test the main elements of the final TT-PET chip.

The ASIC comprises a matrix of 30 square n-on-p pixels of $ 470 \times 470 ~\mathrm{\mu m^2} $ area, with $ 30 ~\mathrm{\mu m} $ inter-pixel spacing. The positive high voltage is applied to the pixels, with a breakdown voltage of approximately $ 210 ~\mathrm{V} $. The depletion region extends for a depth of $ 82 ~\mathrm{\mu m} $ in the $ 1 ~\mathrm{k\Omega cm} $ resistivity substrate, limited by the backside thinning and metallization. This high voltage ensures the generation of an electric field in the sensor bulk above $ 2 ~\mathrm{V/\mu m} $. The large ratio between the pixel side and the sensor thickness is necessary to maximize the sensor uniformity of response in terms of timing \cite{phd_Paolozzi, silicon_tres}. The signal generated in each pixel is routed to the front-end, placed outside the guard ring in the periphery of the chip. 

The front-end comprises a pre-amplifier based on a SiGe HBT transistor and a CMOS-based open-loop tri-stage discriminator. The discrimination threshold can be adjusted independently for each channel with an 8-bit DAC. The output of the discriminator is sent to a fast-OR chain, which preserves the time of arrival and the time over threshold (TOT) of the pixel. The address of the pixel is also registered. Each pixel can be masked, in which case its signal does not propagate to the fast-OR. In this prototype, if two pixels in a chip fire in a single event, only the first time of arrival is registered, while the address of the pixel with the lowest row and column index is assigned to the hit. This simple architecture was chosen for the demonstrator as a robust solution to limit the complexity of the read out logic. In the final chip multiple independent fast-OR lines will be used to handle events with cluster size larger than one.

The time of arrival and the time over threshold of the fast-OR output signal are digitized using a CMOS-based hybrid TDC made of a free-running ring oscillator with a binning of $ 50 ~\mathrm{ps} $ and a $ 700 ~\mathrm{ps} $ counter, developed on purpose for this project \cite{patent_TDC}. Both the counter and the 14 states of the ring oscillator are read for the measurement of the time of arrival and the time over threshold of the signal.

A $ 10 ~\mathrm{MHz} $ clock is distributed to the different chips to offer a common time reference for the TDCs and run the chip logic.

\section{Experimental setup and methods}

\subsection{The experimental setup at the SPS beam test facility at CERN}

\begin{figure}[htbp]
\centering 
\includegraphics[width=.9\textwidth,trim=0 0 0 0]{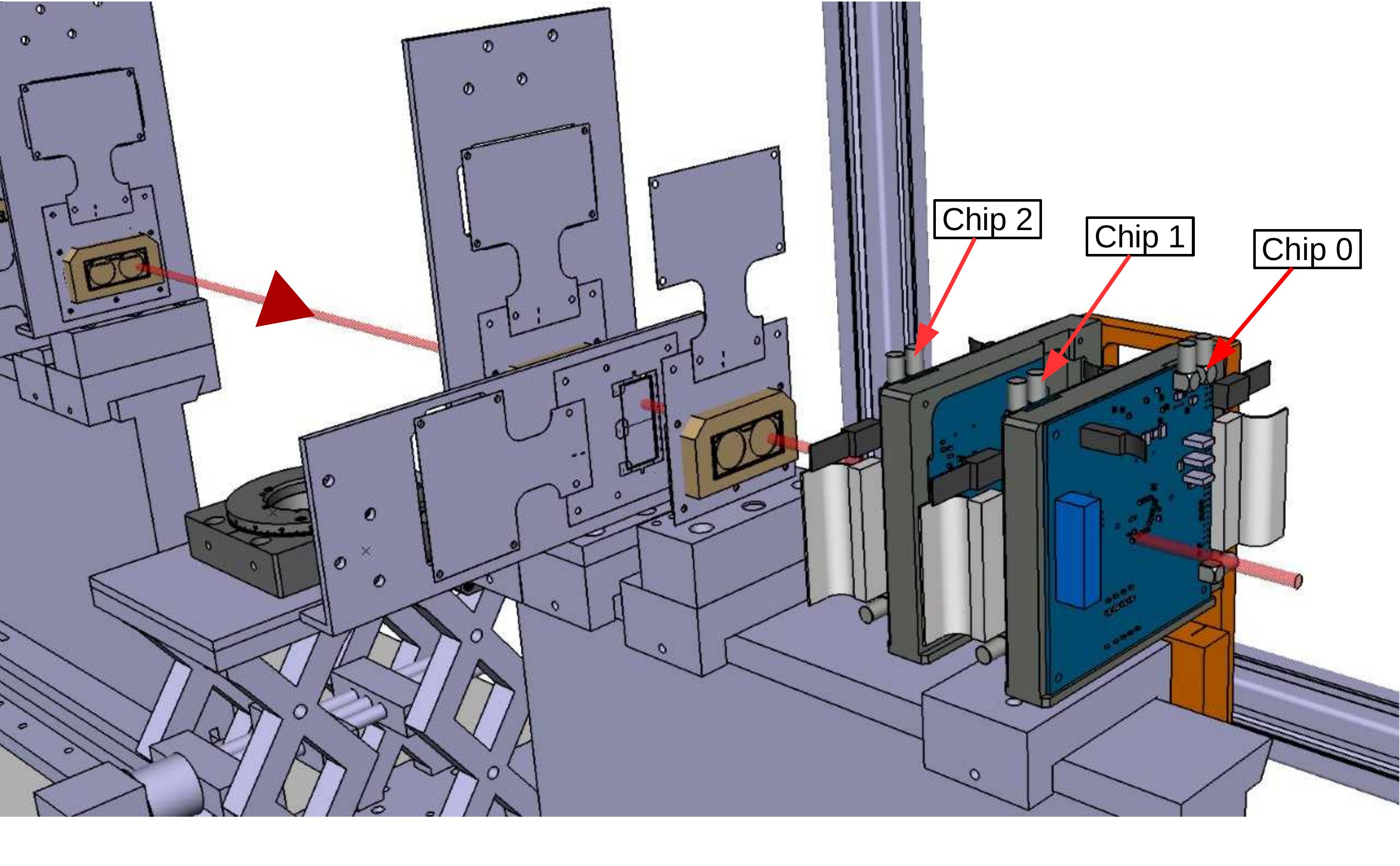}
\caption{\label{fig:setup} The experimental setup at the SPS beam test facility. The red line represents the particle beam. Four of the tracking telescope planes are visible on the left. The three boards with the demonstrator were downstream with respect to the telescope. The board containing chip 0 was rotated by 180 degrees along the vertical axis with respect to the other two boards.}
\end{figure}

In order to study the efficiency, timing performance, front-end noise and uniformity of response, the demonstrator chip was tested at the SPS beam test facility at CERN with MIPs. The experimental setup (Figure \ref{fig:setup}) consisted of a tracking telescope \cite{telescope} that provided the trigger and the particle track parameters to three demonstrator chips. The three chips were read out using a readout system developed at the DPNC, with a custom firmware designed to operate the demonstrator with an external trigger.

The chips were operated at two working points: a low-power working point, with a preamplifier power consumption of $ 160 ~\mathrm{\mu W/channel} $, compliant with the TT-PET power requirements, and, for comparison, a working point with power consumption of $ 375 ~\mathrm{\mu W/channel} $, as was used for the previous monolithic prototype \cite{testbeam_2017}. For both working points a high voltage of $ 180 ~\mathrm{V} $ was applied during data taking.

The nominal threshold was set to $ 15 ~\mathrm{mV} $ above the measured amplifier baseline, corresponding to approximately $ 5 $ standard deviations from the electronic noise for each pixel, with a calibration procedure that minimized the noise hit rate; it was then raised for the efficiency measurement as a function of the threshold. After the calibration, the noise hit rate per chip was measured to be $ 4.3 \times 10^{-3} ~\mathrm{Hz} $ at the nominal threshold. The coincidence time window with the telescope trigger was set to $ 200 ~\mathrm{ns} $, corresponding to a random hit probability per chip in the trigger window lower than $ 10^{-9} $.

During the data taking, the four pixels closer to the I/O pads (corresponding to the three pixels of the rightmost column in figure \ref{fig:p0_layout} and the bottom pixel in the adjacent column) were masked on hardware, due to noise induced on them by the single-ended clock line. The coupling between these pixels and the digital lines was caused by the vicinity of the I/O bump-bonding pads (the octagonal pads inside the red lines to the right side of Figure \ref{fig:p0_layout}), which were not used but still connected to the corresponding signals. These pads will be removed in the final chip and the clock will be distributed using differential lines.

\begin{figure}[htbp]
\centering 
\includegraphics[width=.9\textwidth,trim=0 0 0 0]{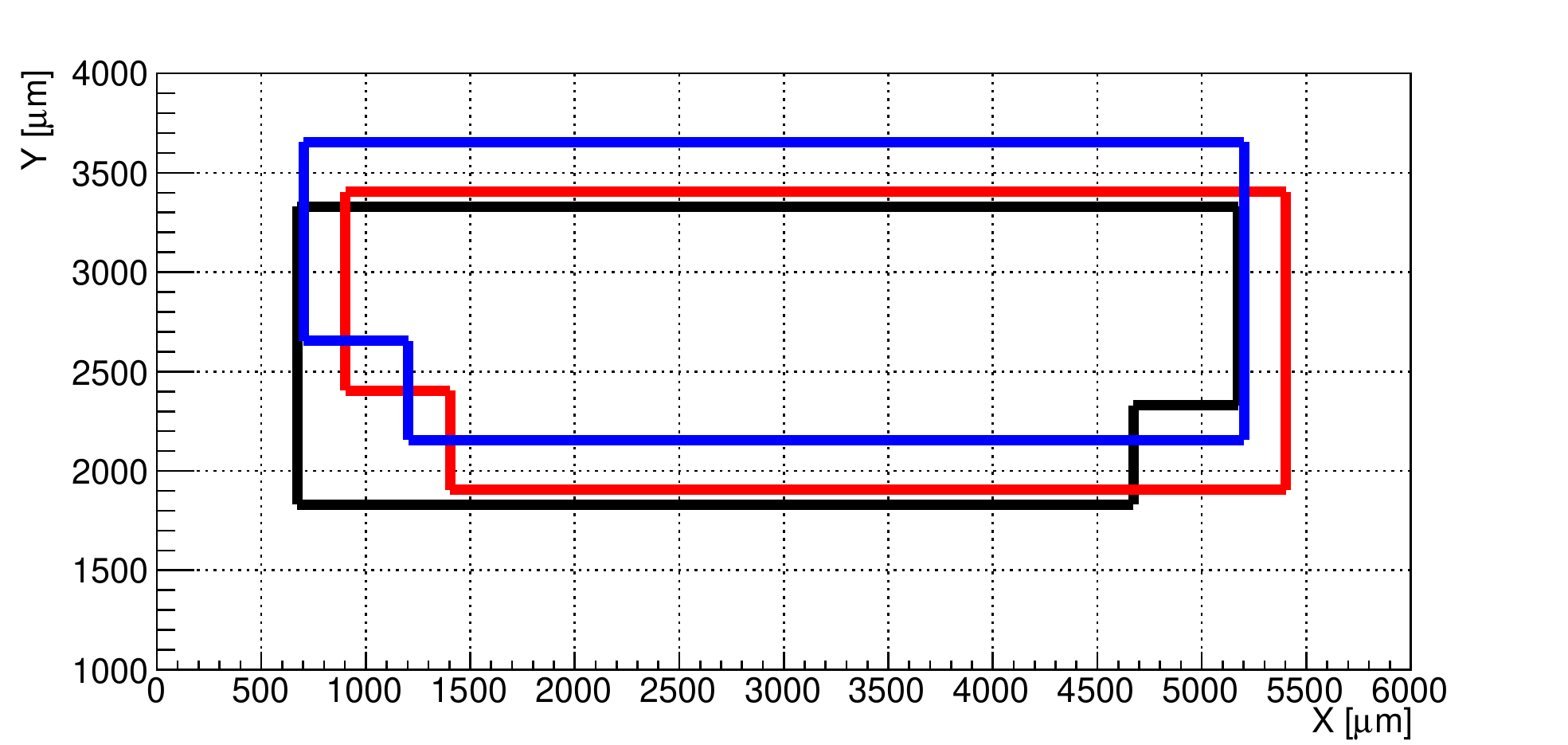}
\caption{\label{fig:chipmap} Relative position of the three demonstrator chips under test in the reference frame of the beam telescope. The active area of the chip 0 is delimited by the black line, of chip 1 by the red line and of chip 2 by the blue line.}
\end{figure}

The relative position of the three chips, measured using the particles of the beam and limited to the active pixels, is shown in figure \ref{fig:chipmap}.

\subsection{Analysis of the data}

To minimize the effects of the multiple scattering and of the tracking-telescope pointing resolution, for the efficiency calculation an area of $ 50 ~\mathrm{\mu m} $ around the edge of the active area of the chip, comprising the region of the pixels closest to the first guard ring, was removed from the analysis. The same exclusion area was used for the calculation of the time resolution.

\begin{figure}[htbp]
\centering 
\includegraphics[width=.9\textwidth]{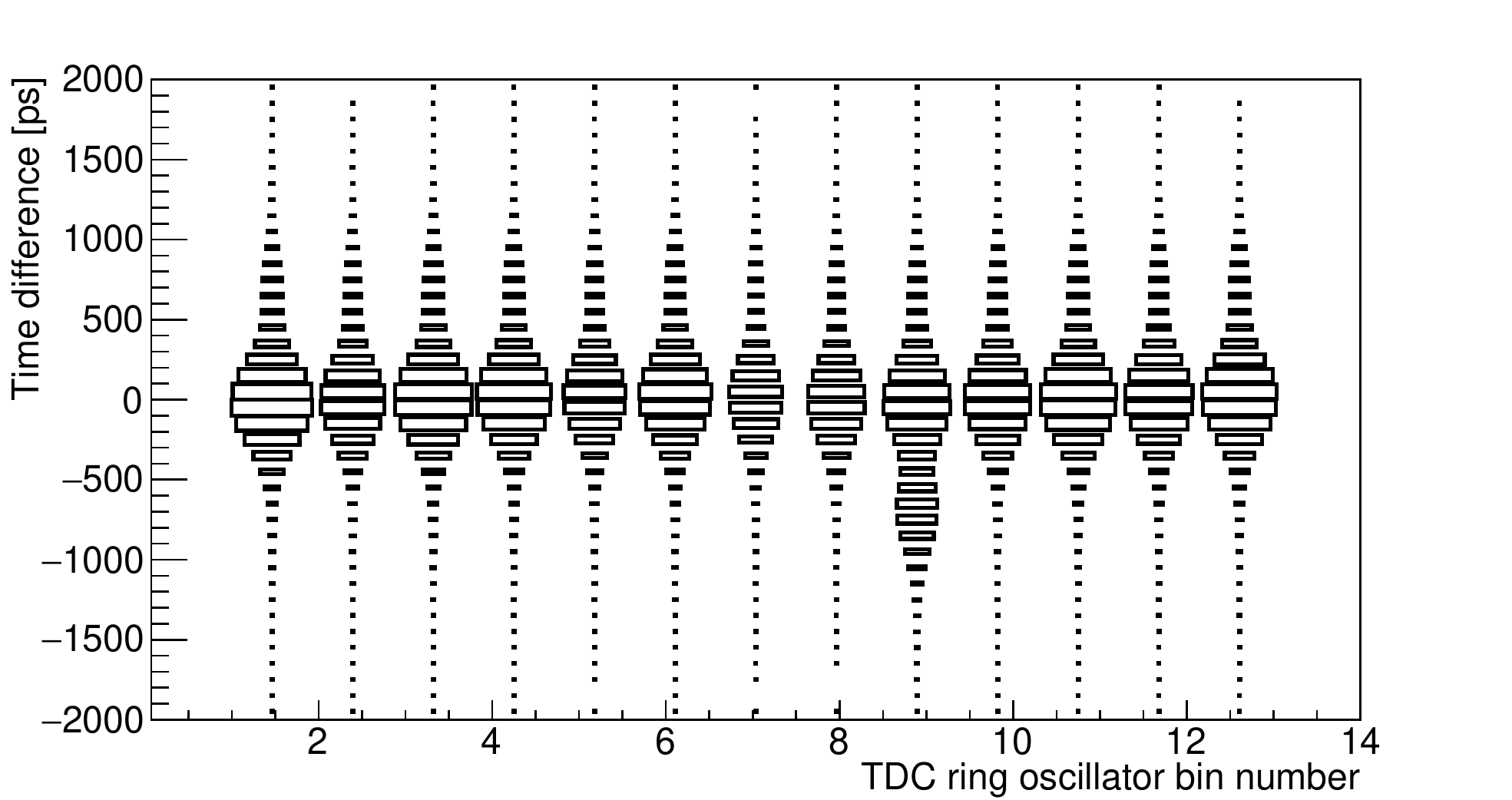}
\caption{\label{fig:TDC_error} Time difference between chip 0 and chip 1 as a function of the status measured by the TDC ring oscillator. The counter starting-time uncertainty, visible for bin 9, causes a systematic error of one period of the ring oscillator, corresponding to $ 700 ~\mathrm{ps} $.}
\end{figure}

A cut was applied during the analysis to remove the effect of a small issue, that has been identified thanks to this prototype: due to a minor design flaw of the TDC, the counter had an uncertainty of 1 bit for the events recorded in one of the fourteen states of the TDC ring oscillator. For this reason, the events in bin 9 of the TDC ring oscillator in Figure \ref{fig:TDC_error} were removed from the analysis. This selection reduces the available statistics without introducing a bias on the data sample, since the time of arrival of the hit is asynchronous with respect to the phase of the TDC ring oscillator. This error will be corrected in the final chip design.

\section{Results}

\subsection{Efficiency and front-end noise}

The efficiency of the three chips under test was measured for the two amplifier power consumption working points. The results obtained at the nominal threshold are reported in table \ref{tab:efftable}. When calculating the efficiency, to reduce further the effect of the multiple scattering and of the beam-telescope resolution near the chip border, a hit confirmation was requested on the other two chips under test.

\begin{table}[htbp]                                                                                                                                                                                                 
\centering                                                                                                                                                                                                          
\caption{\label{tab:efftable} Efficiency at nominal threshold for the three demonstrator chips under test for different values of the amplifier power consumption. Only the statistical error is reported. The two values of the power consumption correspond to the working point of the front-end for the TT-PET scanner ($ 160 ~\mathrm{\mu W/ch} $) and the working point used for testing the previous prototype \cite{testbeam_2017} ($ 375 ~\mathrm{\mu W/ch} $).}                                                   
\smallskip                                                                                
\begin{tabular}{c|ccc|}                                                                   
\cline{1-4}
\multicolumn{1}{ |c| } {Power consumption} & \multicolumn{3}{ c| }{Efficiency [$ \% $]}\\
\cline{2-4}
\multicolumn{1}{ |c| } {[$ \mathrm{\mu W/ch} $]} & Chip 0 & Chip 1 & Chip 2 \\
\hline
\multicolumn{1}{ |c| } { $ 160 $ } & $ 99.933 \pm 0.007 $ & $ 99.986 \pm 0.003 $ & $ 99.985 \pm 0.003 $ \\
\multicolumn{1}{ |c| } { $ 375 $ } & $ 99.924 \pm 0.006 $ & $ 99.980 \pm 0.003 $ & $ 99.973 \pm 0.003 $ \\
\hline                                                                                    
\end{tabular}
\end{table}

\begin{figure}[htbp]
\centering 
\includegraphics[width=.99\textwidth]{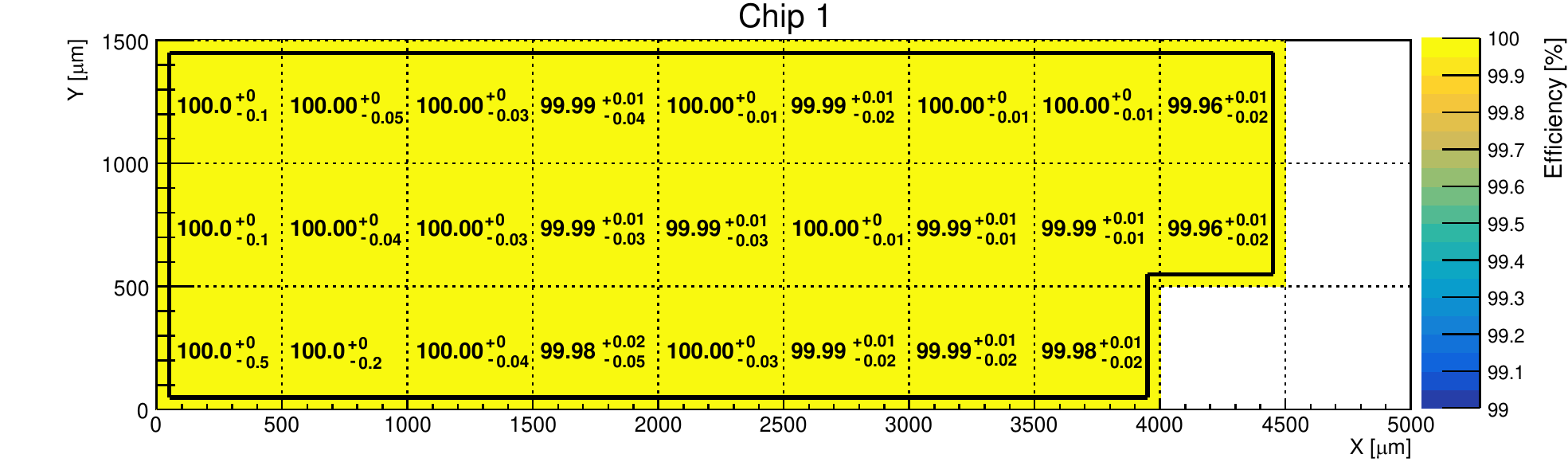}
\caption{\label{fig:effmap} Pixel efficiency map of chip 1. The colored area is the chip active area. The dashed lines represent the separation between pixels. The continuous line represents the border of the area used for the efficiency calculation. The four pixels in white were masked in hardware as discussed in the text.}
\end{figure}

Figure \ref{fig:effmap} shows the pixel efficiency for the chip in the center (chip 1), measured at the nominal threshold value for the low-power working point. The line in the figure indicates the borders of the area used to calculate the efficiency.

\begin{figure}[htbp]
\centering
\includegraphics[width=.6\textwidth]{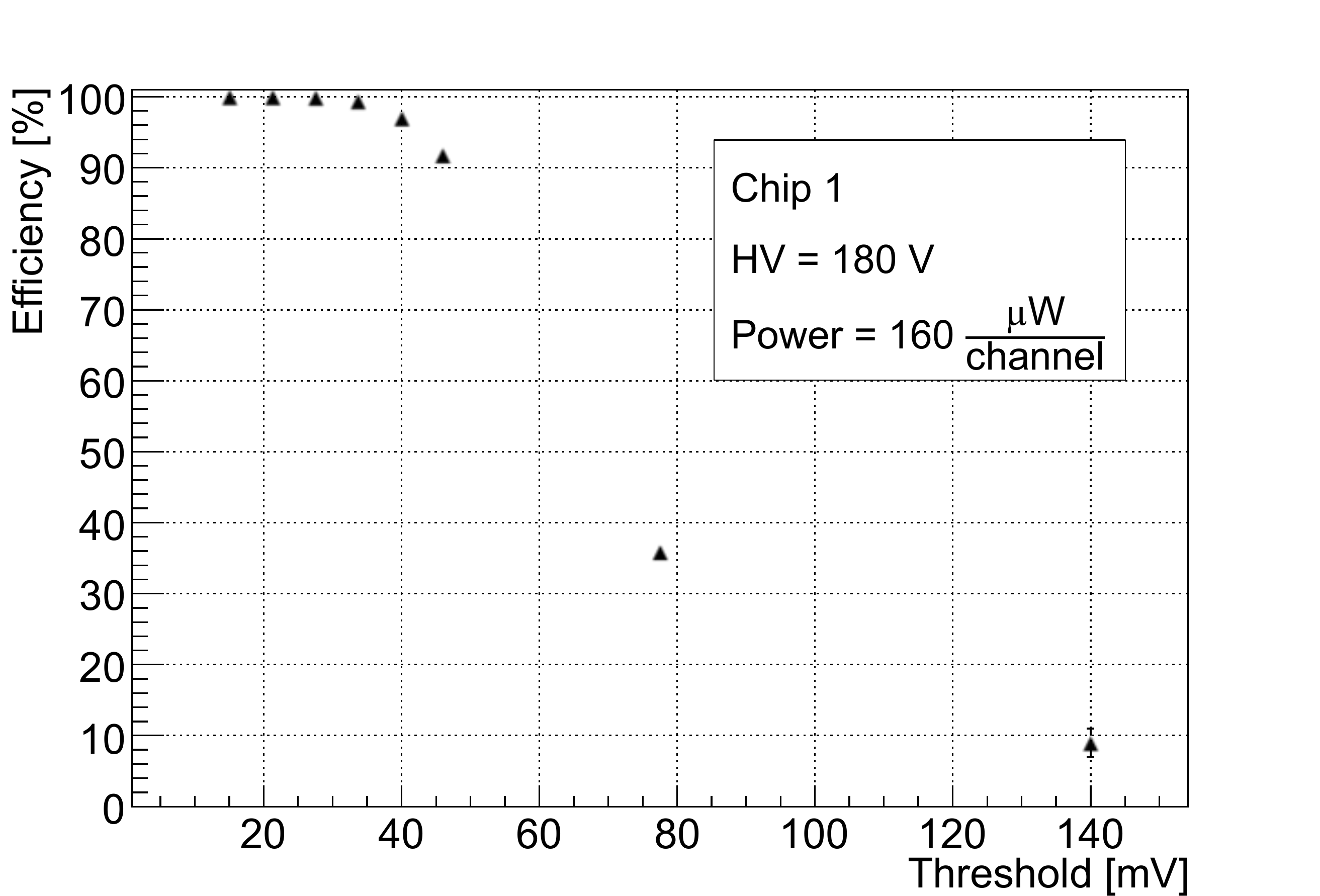}
\caption{\label{fig:effscan} Efficiency as a function of the voltage threshold for chip 1. The voltage threshold value shown in the horizontal axis represents the difference between the global threshold setting and the typical voltage offset at the output of the preamplifier. The mismatch between different channels is corrected with the pixel DAC. For the nominal working point the threshold was set to $ 15 ~\mathrm{mV} $ above the amplifier baseline.}
\end{figure}

Figure \ref{fig:effscan} shows the efficiency of chip 1 measured at different thresholds at the low-power operating point. The data show the noise margin for sensor operation, with the efficiency plateau extending over a factor two above the nominal discriminator threshold. These data, obtained with minimum ionizing particles crossing perpendicularly the sensor, are compatible with an amplifier gain of $ (50\pm5) ~\mathrm{mV/fC} $. 

The equivalent noise charge (ENC) of the front-end was estimated from the gain and noise rate measured at the nominal threshold\footnote{This estimation of the ENC using the discrimination threshold is affected by the discriminator response, that shows a non-linearity for the smallest signals, effectively filtering part of the amplifier noise and decreasing the nominal threshold. The ENC of the amplifier, that contributes to the detector time resolution, is expected to be higher, as calculated in \cite{pierpaolo_chip}.}. The lowest threshold corresponds to at least $ 5 $ standard deviations of the amplifier voltage noise (compatible with such a small noise hit rate and in accordance to bench measurements described in \cite{pierpaolo_chip}). The front-end ENC can be estimated as:

\begin{equation*}
ENC = \frac{\frac{Vth_{min}}{N_{\sigma_{noise}}}}{Gain} ~\approx ~350~e^{-}
\end{equation*}
where $ N_{\sigma_{noise}}=5 $ is the number of standard deviations of the voltage noise corresponding to the nominal threshold. Therefore the nominal threshold of $ 15 ~\mathrm{mV} $ corresponds approximately to $ 1750 ~\mathrm{e^{-}} $.

\subsection{Time resolution}

\begin{figure}[htbp]
\centering 
\includegraphics[width=.51\textwidth, trim=0 15 0 20]{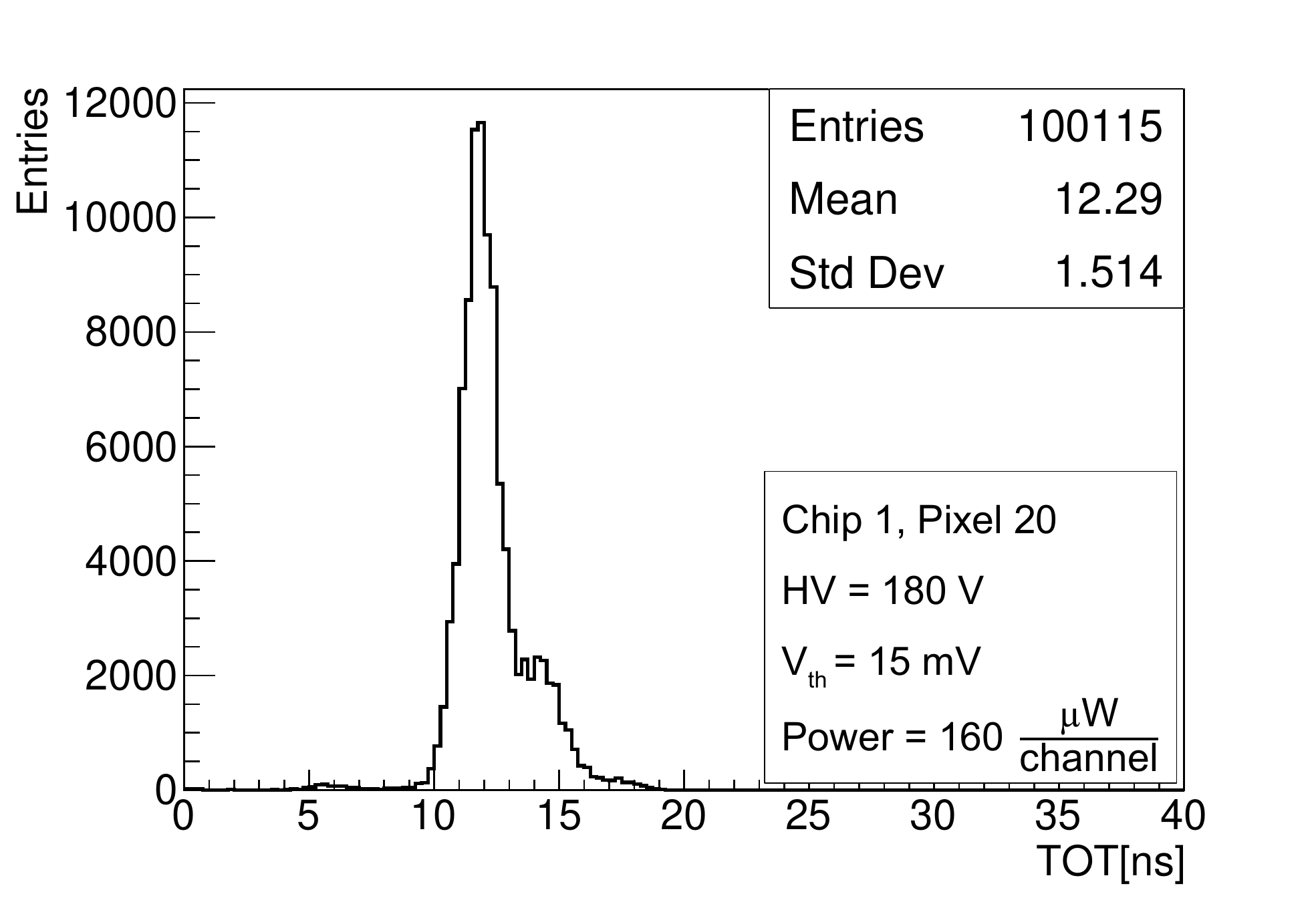}
\qquad
\includegraphics[width=.42\textwidth, trim=80 0 0 20]{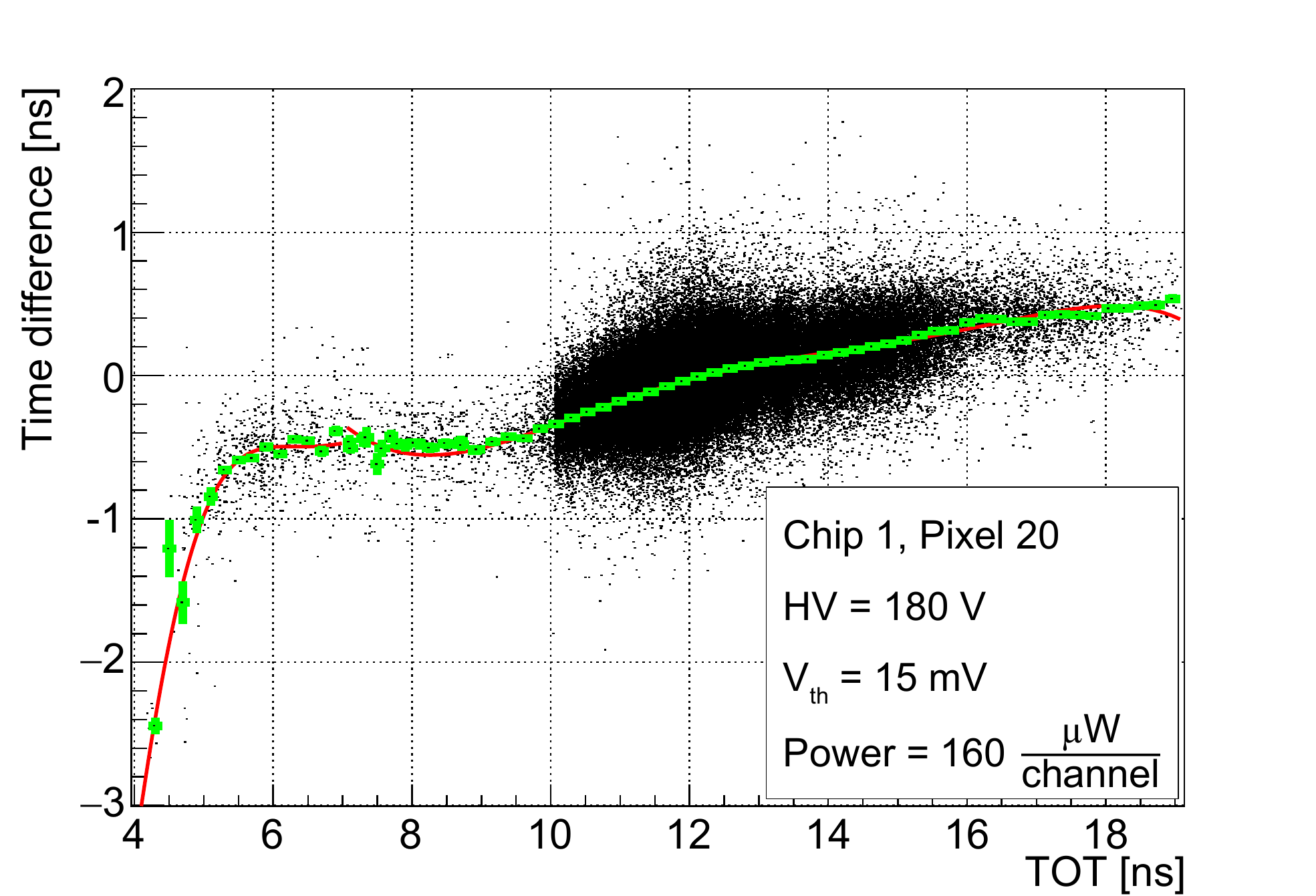}
\caption{\label{fig:tottw} (Left) Signal time over threshold distribution for one of the pixels of chip 1. (Right) Example of time-walk correction function for one of the pixels of chip 1. The red lines are the result of a polynomial fit of the mean values of the time difference for TOT bins of $ 0.25 ~\mathrm{ns} $, represented by the green dots.}
\end{figure}

Figure \ref{fig:tottw} left shows the distribution of the signal TOT for the hits recorded by one of the pixels of chip 1 at the low-power working point. Different TOT peaks are visible in the figure and they can be attributed to different reasons. We attribute the first peak, visible between $ 3 ~\mathrm{ns} $ and $ 7 ~\mathrm{ns} $, to a non linear response of the discriminator for the smallest signals. The main peak, at $ 12 ~\mathrm{ns} $, corresponds to the most probable charge deposited into the sensor by a MIP. It is followed by secondary peaks at $ 14 ~\mathrm{ns} $ and $ 17 ~\mathrm{ns} $. A possible explanation for these peaks is a small residual noise induced by the single-ended digital trigger signal, affecting the grounding of the pixel matrix. In this scenario, the time difference between the peaks would be caused by the delay of the fast-OR line. The first peak between $ 3 $ and $ 7 ~\mathrm{ns} $, visible with MIPs, should not be present when operating the sensor with more ionizing radiation, as in the case of the TT-PET scanner. The digital cross-talk, on the contrary, can be reduced only improving the design at system level. The introduction of slower trigger signals in a differential configuration in the final chip will eliminate this problem.

For the calibration and measurement of the time resolution, hits with a TOT between $ 2 ~\mathrm{ns} $ and $ 40 ~\mathrm{ns} $ were selected. The time calibration consisted of two steps: the time-walk correction and the time-skew correction. As an example, Figure \ref{fig:tottw} right shows the time difference between chip 1 and chip 0 as a function of the TOT of chip 1, for one of the pixels of chip 1. These distributions were used for the time-walk correction. The green segments represent the mean value of the time difference for TOT slices of $ 0.25 ~\mathrm{ns} $. These mean values were fitted with two polynomial curves. The use of different curves for different TOT intervals is motivated by the non linear response of the discriminator for the smallest signals. The range of the time-walk correction spans approximately $ 1 ~\mathrm{ns} $, making this calibration fundamental to obtain a $ 100 ~\mathrm{ps} $ time resolution. 

The time skew between different pixels of the same chip is mostly generated by the different path of the signals in the fast-OR line. To correct for this effect, the average time of arrival, corrected for time-walk, is set to the same value for each pixel. The initial time-skew between two pixels was measured to be as large as $ 2 ~\mathrm{ns} $ and is affected by the mismatch of the electronic components.

\begin{figure}[htbp]
\centering 
\includegraphics[width=.45\textwidth, trim=20 0 50 0]{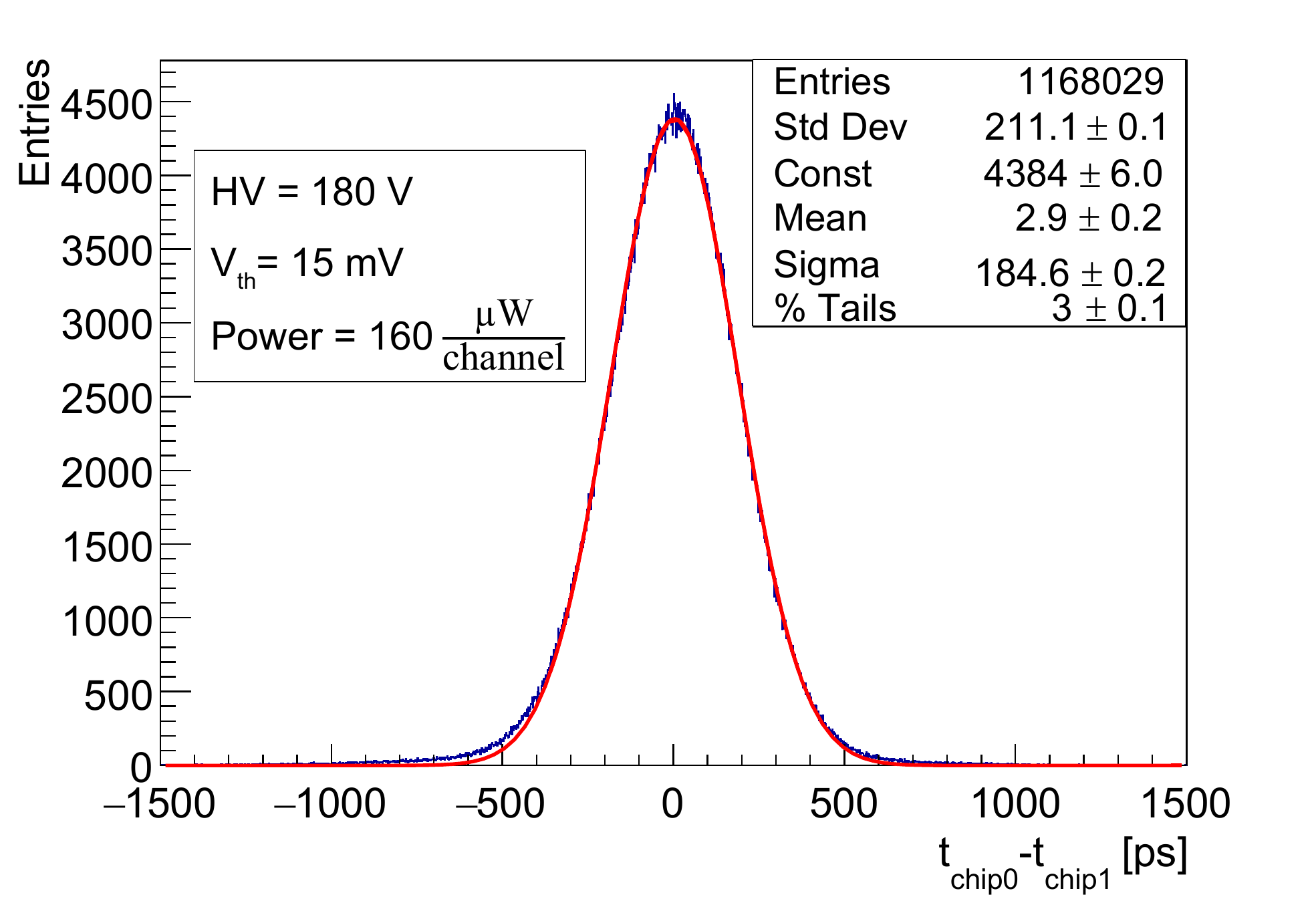}
\qquad
\includegraphics[width=.46\textwidth, trim=50 0 20 10]{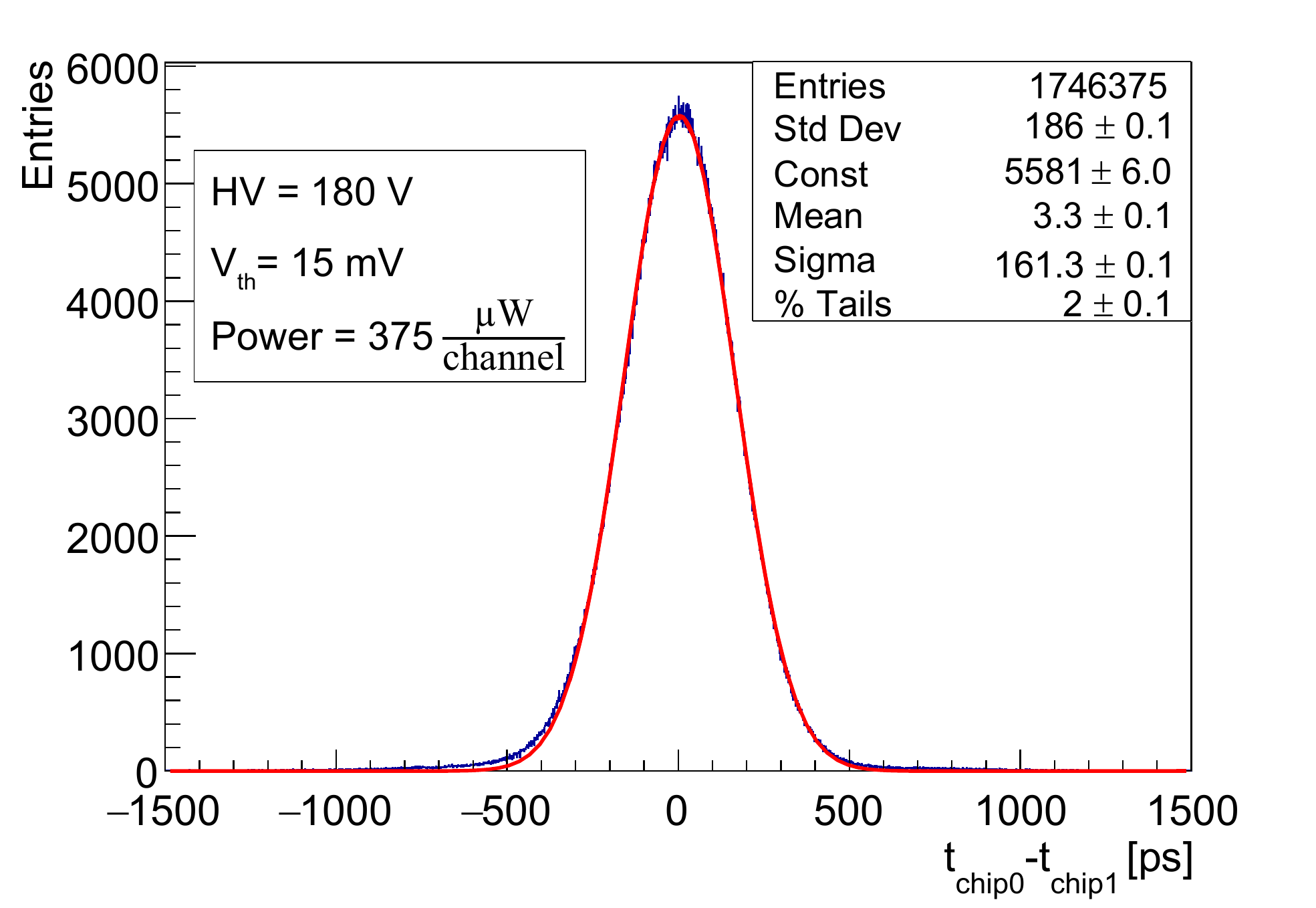}
\caption{\label{fig:tres} Time difference between chip 0 and chip 1 for an amplifier power consumption of $ 160 ~\mathrm{\mu W/channel} $ (left) and  $ 375 ~\mathrm{\mu W/channel} $ (right).}
\end{figure}

The time resolution can be obtained by measuring the jitter of the time of flight between two calibrated chips. Figure \ref{fig:tres} shows the difference of the time of arrival between chip 0 and chip 1 for the low-power (left) and high-power (right) working point. The mean value, not significant for this study, was set to zero. The core of the distributions was fitted with a gaussian function in the $ \pm ~2 $ standard deviations interval. The non-gaussian behavior of the tails was then measured as the fraction of events exceeding the gaussian functions in the entire range. The non-gaussian part of the tails was found to be a few percent.

Table \ref{tab:trestable} shows the resolution of the time of flight for the combinations of the three chips at the two working points as well as the corresponding single-detector time resolution. The three detectors show similar timing performance. 

\begin{table}

\centering
\caption{\label{tab:trestable} (Left) Resolution of the time of flight between the three combinations of the chips under test. The $ \mathrm{\sigma} $ is the standard deviation of the gaussian fit to the $ \pm 2 $ standard deviations core of the time-of-flight distribution. (Right) Time resolution of the three chips under test, obtained from the measurement of the time-of-flight resolution.}
\smallskip
\begin{tabular}{c|cc|}
\cline{2-3}
& \multicolumn{2}{ c| }{TOF resolution [$ \mathrm{ps} $]}\\
\cline{2-3}
& \multicolumn{1}{ |c }{low-power} &  \multicolumn{1}{ c| }{high-power} \\
\hline
\multicolumn{1}{ |c| } {$ \sigma_{TOF,0-1} $} & $ 184.6 \pm 0.2 $ & $ 161.3 \pm 0.1 $ \\
\multicolumn{1}{ |c| } {$ \sigma_{TOF,0-2} $} & $ 180.0 \pm 0.2 $ & $ 157.3 \pm 0.1 $ \\
\multicolumn{1}{ |c| } {$ \sigma_{TOF,1-2} $} & $ 184.9 \pm 0.2 $ & $ 161.2 \pm 0.1 $ \\
\hline
\end{tabular}
\quad
\smallskip
\begin{tabular}{c|cc|}
\cline{2-3}
& \multicolumn{2}{ c| }{Time resolution [$ \mathrm{ps} $]}\\
\cline{2-3}
& \multicolumn{1}{ |c }{low-power} &  \multicolumn{1}{ c| }{high-power} \\
\hline
\multicolumn{1}{ |c| } { \bfseries\mathversion{bold} $ \sigma_{t, ~chip ~0} $ } & \bfseries 127.3 $ \pm $ 0.2 & \bfseries 111.3 $ \pm $ 0.1 \\
\multicolumn{1}{ |c| } { \bfseries\mathversion{bold} $ \sigma_{t, ~chip ~1} $ } & \bfseries 134.2 $ \pm $ 0.2 & \bfseries 116.7 $ \pm $ 0.1 \\
\multicolumn{1}{ |c| } { \bfseries\mathversion{bold} $ \sigma_{t, ~chip ~2} $ } & \bfseries 127.2 $ \pm $ 0.2 & \bfseries 111.2 $ \pm $ 0.1 \\
\hline
\end{tabular}
\end{table}

\begin{figure}[htbp]
\centering %
\includegraphics[width=.99\textwidth]{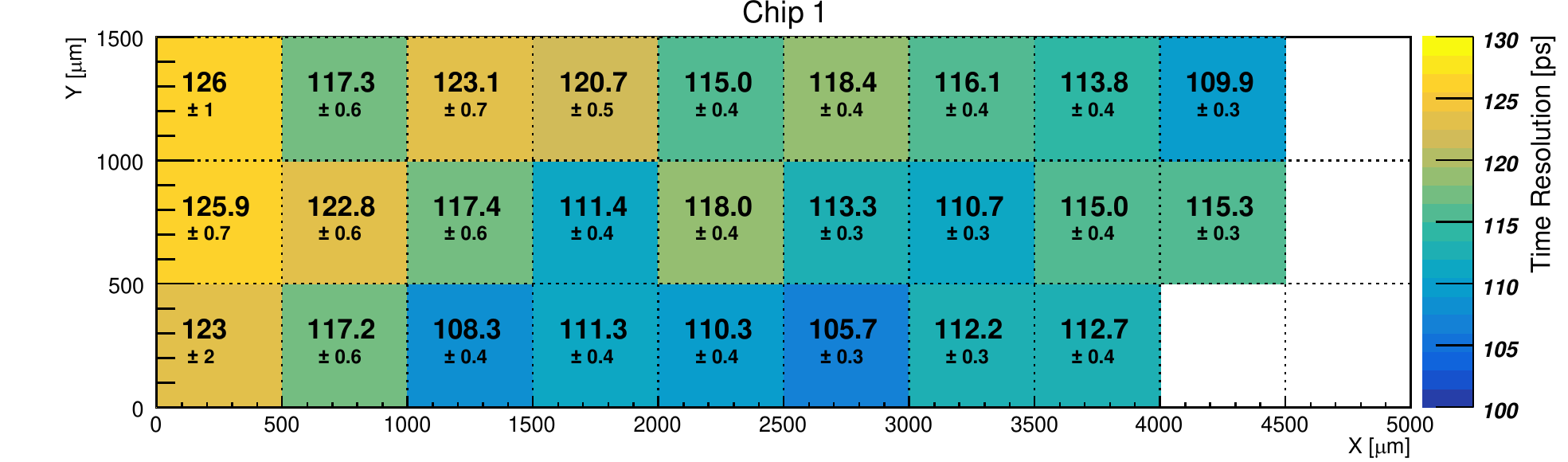}
\caption{\label{fig:tresmap} Time resolution of the pixels of chip 1. The dashed lines represent the separation between different pixels. The map shows the pixel matrix oriented as in Figure \ref{fig:p0_layout}. The error is statistical only.}
\end{figure}

To give an idea of the uniformity of response within a chip, Figure \ref{fig:tresmap} shows the map of the time resolution of the pixels of chip 1 for the high-power working point. The map was obtained selecting the tracks that pointed to each pixel and measuring the time resolution using chip 0 as reference. To obtain the contribution from chip 1, the jitter of the time of flight was divided by $ \sqrt{2} $. A steady small worsening of the time resolution towards the left of the map is visible. An hypothesis to explain this effect is the larger impedance of the ground line for the front-end channels far from the chip ground connection that is done in the right side of the chip.

\section{Conclusions}

The demonstrator of the fast, monolithic ASIC of the TT-PET project was produced and tested with minimum ionizing particles. The biasing structures, the pixel matrix, the fast-OR line and the TDC were qualified and the minor modifications required for the final chip design were identified. The measurements, done at a low-power ($ 160 ~\mathrm{\mu W/channel} $) and a high-power ($ 375 ~\mathrm{\mu W/channel} $) working point, show an efficiency above $ 99.9 ~\mathrm{\%} $ when the chip was operated at the nominal threshold of $ 15 ~\mathrm{mV} $, with a noise hit rate per chip of $ 0.004 ~\mathrm{Hz} $. The front-end noise, estimated from the efficiency measurement, is $ 350 ~\mathrm{e^-} ~\mathrm{RMS} $. At the low-power working point, compatible with the power-budget of the TT-PET scanner, the time resolution was measured to be $ 130 ~\mathrm{ps} ~\mathrm{RMS} $. A time resolution as low as $ 110 ~\mathrm{ps} ~\mathrm{RMS} $ was measured at the high-power working point, showing an improvement of a factor 2 with respect to the results of the first prototype of the TT-PET chip.

\acknowledgments

The authors wish to thank the technical staff of the DPNC of the University of Geneva for the design and construction of the instrumentation that made this test possible and the colleagues of the University of Geneva who operated the test beam telescope and reconstructed the particle track data. Finally, the authors wish to thank the Swiss National Science Foundation, which supported this research with the SINERGIA grant CRSII2\_160808.



\end{document}